\documentstyle[12pt]{article}
\begin{document}
\rightline{Jan 1994}
\rightline{McGill/94-10}

\begin{center}

{\Large \bf $\mbox{SU}(3)_L \otimes \mbox{U}(1)_N$ and
$\mbox{SU}(4)_L \otimes \mbox{U}(1)_N$
gauge models with right-handed neutrinos}\\
\vspace{2cm}

Robert Foot\footnote{Email: Foot@hep.physics.mcgill.ca}\\
Physics Department, McGill University\\ 3600 University Street, Montr\'eal,
Qu\'ebec Canada H3A 2T8 \\
\vspace{0.5cm}

Hoang Ngoc Long\\
Institute of Theoretical Physics\\ National Center for Natural Science and
Technology\\ P.O.Box 429, Boho, Hanoi 10000, Vietnam\\
\vspace{0.5cm}

and\\
\vspace{0.5cm}

Tuan A.Tran\\
Graduate College in Physics\\  National Center for Natural Science and
Technology\\ P.O.Box 429, Boho, Hanoi 10000, Vietnam\\
\vspace{1cm}

{\bf Abstract}\\
\end{center}

Pisano and Pleitez have introduced an interesting $\mbox{SU}(3)_C
\otimes \mbox{SU}(3)_L \otimes \mbox{U}(1)_N$
gauge model which has the property that gauge anomaly cancellation
requires the number of generations to be a multiple of 3.
We consider generalizing that model to incorporate
right-handed neutrinos. We find that there exists a non-trivial
generalization of the Pisano-Pleitez model with right-handed neutrinos
which is actually simpler than the original model in that symmetry breaking
can be achieved with just three $\mbox{SU}(3)_L$ triplets (rather than
3 $\mbox{SU}(3)_L$ triplets and a sextet). We also consider
a gauge model based on $\mbox{SU}(3)_C\otimes \mbox{SU}(4)_L
\otimes \mbox{U}(1)_N$ symmetry.
Both of these new models also have the feature that
the anomalies cancel only when
the number of generations is divisible by 3.\\

PACS number(s): 12.15.Ce, 12.15.Ff\\

\newpage

The standard model of electroweak interactions and QCD has proven
extremely successful. In the last decade no new elementary particles
or interactions have been discovered beyond the expected confirmation
of the $W$ and $Z$ bosons of the standard model. To go beyond the
standard model, we should look for interesting ideas until novel
events are discovered.

Pisano and Pleitez [1] have proposed an interesting model based
on the gauge group
\begin{equation} \mbox{SU}(3)_c \otimes \mbox{SU}(3)_L \otimes
\mbox{U}(1)_N \end{equation}
(for further work on this model see Ref. [2,3].)
This model has the interesting feature that each generation of
fermions is anomalous, but that with three generations the
anomalies cancelled.

In this paper we point out that if right-handed neutrinos are
included then there is an alternative $\mbox{SU}(3)_L \otimes \mbox{U}(1)_N$
gauge model. This alternative model is actually simpler than
the Pisano Pleitez model because it turns out that
less Higgs mulitplets are needed
in-order to allow the fermions to gain masses and to
break the gauge symmetry. This alternative model also
has the interesting property that
anomalies only cancel when all three generations are included.
This alternative model allows for Dirac neutrino masses.
We will also discuss a $\mbox{SU}(3)_C \otimes \mbox{SU}(4)_L \otimes
\mbox{U}(1)_N$
gauge model which also includes right-handed neutrinos and
also requires 3 generations to cancel the anomalies.

We start by briefly reviewing the Pisano-Pleitez model.
In that model the three lepton generations transform under the
gauge symmetry, Eq.(1) as
\begin{equation}f^a_L = \left(\begin{array}{c}
\nu_L^a\\
e_L^a\\
(e_R^c)^a
\end{array}\right) \sim ( 1, 3, 0), \end{equation}
where $a = 1, 2, 3$ is the generation index.

Two of the three quark generations transform identically and one
generation, transforms in a different representation of $\mbox{SU}(3)_C \otimes
\mbox{SU}(3)_L \otimes \mbox{U}(1)_N$:
\begin{equation}\begin{array}{l}
Q_{1L} = \left(\begin{array}{c}
u_1\\
d_1\\
J_1\end{array}\right)_L \sim (3, 3, 2/3),\\
u_{1R} \sim (3, 1, 2/3),\
d_{1R} \sim (3, 1, -1/3),\ J_{1R} \sim (3, 1, 5/3),\\
Q_{iL} = \left(\begin{array}{c}
d_i\\
u_i\\
J_i\end{array}\right)_L \sim (3, \bar 3, -1/3), \\
u_{iR} \sim (3, 1, 2/3),\ d_{iR} \sim (3, 1, -1/3),
J_{iR} \sim (3, 1, -4/3),\end{array} \end{equation}
where $i = 2,3$.

Symmetry breaking and fermion mass generation can be achieved by
three scalar $\mbox{SU}(3)_L$ triplets and a sextet. For these details
the reader can see Ref.[1].

If right-handed neutrinos are included then we can add either three
$\mbox{SU}(3)_L \otimes \mbox{U}(1)_N$ singlets
$\nu_R^a \sim (1, 1, 0)$ (which is a
trivial generalization of the Pisano-Pleitez model), or we can try
to modify the quantum numbers of the fermions such that $\nu_R$
replaces $e_R$ as the third component of the lepton triplets.
We will show that this latter case is possible and it leads to a
$\mbox{SU}(3)_L$ model which is simpler than the Pisano-Pleitez model with
the same nice features.

The gauge quantum numbers of the fermions are as follows:
The leptons consist of:
\begin{equation}f^a_L = \left(\begin{array}{c}
\nu_L^a\\
e_L^a\\
(\nu_R^c)^a
\end{array}\right) \sim (1, 3, -1/3), \  e_R^a \sim (1, 1, -1),
\end{equation}
where $a = 1,...,3$. As in the case of the Pisano-Pleitez model,
two of the quark generations transform identically, and one generation
transforms differently:
\begin{equation}\begin{array}{l}
Q_{1L} = \left(\begin{array}{c}
u_{1L}\\
d_{1L}\\
u^{'}_{1L} \end{array}\right) \sim (3, 3, 1/3),\\
u_{1R} \sim (3, 1, 2/3),\   d_{1R} \sim (3, 1, -1/3), \
u_{1R}^{'} \sim (3, 1, 2/3),\\
Q_{iL} = \left(\begin{array}{c}
d_{iL}\\
u_{iL}\\
d^{'}_{iL}\end{array}\right) \sim (3, \bar 3, 0),\\
u_{iR} \sim (3, 1, 2/3), \  d_{iR} \sim (3, 1, -1/3),\
d_{iR}^{'} \sim (3, 1, -1/3), \end{array}\end{equation}
where $i= 2,3$. It is straightforward to check that all gauge anomalies
cancel with the above choice of gauge quantum numbers. As in the Pisano-Pleitez
model, each generation is anomalous but with all three generations the
anomalies cancel. Symmetry breaking and fermion mass generation can
be achieved with just three $\mbox{SU}(3)_L$ Higgs triplets. We define
them by their Yukawa Lagrangians as follows:
\begin{equation}{\cal L}^{\chi}_{yuk}= \lambda_1 \bar Q_{1L} u^{'}_{1R} \chi
+ \lambda_{2ij} \bar Q_{iL} d^{'}_{jR} \chi^{*} + H.c., \end{equation}
where $\chi \sim (1, 3, -1/3)$ and if
$\chi$ gets the vacuum expectation value (VEV):
\begin{equation}
\langle \chi \rangle = \left(\begin{array}{c}
0\\
0\\
w\end{array}\right), \end{equation}
then the exotic $2/3$ and $-1/3$ quarks gain masses and the
gauge symmetry is broken to the standard model gauge symmetry:
\begin{equation}
\begin{array}{c}
SU(3)_c \otimes SU(3)_L \otimes U(1)_N \\
\downarrow \langle \chi \rangle\\
SU(3)_c \otimes SU(2)_L \otimes U(1)_Y\end{array}
\end{equation}
where $Y = 2N - \sqrt{3}\lambda_8/3$ ($\lambda_8 = diag(1,1,-2)/\sqrt{3}$),
is the combination of $N$ and $\lambda_8$ which annihilates the VEV
(i.e. $Y\langle \chi \rangle = 0$). Note that $Y$ is identical
to the standard hypercharge of the standard model.
Electroweak symmetry breaking and ordinary fermion mass generation are achieved
with two $\mbox{SU}(3)_L$ triplets $\rho, \eta$ which we define through
their Yukawa Lagrangians as follows:
\[ {\cal L}_{Yuk}^{\rho} = \lambda_{1a} \bar Q_{1L} d_{aR} \rho
+ \lambda_{2ia} \bar Q_{iL} u_{aR} \rho^{*} + G_{ab}
\bar f^a_L(f^b_L)^c \rho^{*} + G^{'}_{ab} \bar f_L^a e_R^b \rho + H.c., \]
\[{\cal L}^{\eta}_{yuk} = \lambda_{3a} \bar Q_{1L} u_{aR} \eta +
\lambda_{4ia} \bar Q_{iL} d_{aR} \eta^{*} + H.c.,\]
\begin{equation} \end{equation}
where $\rho \sim (1, 3, 2/3), \ \eta \sim (1, 3, -1/3)$ and we
require the vacuum structure:
\begin{equation}
\langle \rho \rangle = \left(\begin{array}{c}
0\\
u\\
0\end{array}\right), \
\langle \eta \rangle = \left(\begin{array}{c}
v\\
0\\
0\end{array}\right).\end{equation}
The VEV $\langle \rho \rangle$ will generate masses for the three charged
leptons and two of the neutrinos will gain degenerate Dirac masses (with
one necessarily massless) and two up-type quarks and one
down-type quark will also gain masses while the
VEV $\langle \eta \rangle$ will generate masses
for the remaining quarks. The VEVS $\langle \rho \rangle, \langle \eta \rangle$
also give the electroweak gauge bosons masses and results in the
symmetry breaking:
\begin{equation}
\begin{array}{c}
SU(3)_c \otimes SU(3)_L \otimes U(1)_N\\
\downarrow \langle \chi \rangle\\
SU(3)_c \otimes SU(2)_L \otimes U(1)_Y\\
\downarrow \langle \rho \rangle, \langle \eta \rangle \\
SU(3)_c \otimes U(1)_Q \end{array} \end{equation}
Note that because $\langle \rho \rangle, \langle \eta \rangle$
transform as part of a $Y =1,\ \mbox{SU}(2)_L$ doublet
(under the $\mbox{SU}(2)_L \otimes \mbox{U}(1)_Y$
subgroup of $\mbox{SU}(3)_L \otimes \mbox{U}(1)_N$) the
correct $W, Z$ mass relation
ensues, and the model essentially reduces to the standard model
provided $\langle \chi \rangle \gg \langle \eta \rangle, \langle \rho \rangle$.

An important phenomenological difference between this model
and the Pisano-Pleitez
model is that the exotic quarks have electric charges $2/3$ and $-1/3$.
(in the Pisano-Pleitez model the exotic quarks had electric
charges $5/3$ and $-4/3$).
A consequence of this is that the exotic quarks can mix with the ordinary ones.
Indeed, in eq.(9), we can have extra terms obtained by replacing
$d_{aR}, u_{aR}$ with $d^{'}_{iR}, u^{'}_{1R}$. One important consequence
of this type of mixing is that small Flavour changing neutral currents
(FCNCs) will be induced due to the breakdown of the GIM mechanism (one
can easily check however that as $\langle \chi \rangle$ goes to infinity
these induced FCNCs go to zero). This type of situation has been
discussed previously and bounds on the mixing strengths can be
obtained from the experimental non-observation of FCNCs beyond those
predicted by the standard model [4].

We now turn to another possibility beyond the standard model. This
extension of the standard model is based on
$\mbox{SU}(3)_C\otimes \mbox{SU}(4)_L\otimes \mbox{U}(1)_N$ gauge group [3].
In the $\mbox{SU}(3)_C\otimes \mbox{SU}(4)_L\otimes \mbox{U}(1)_N$ model,
the three lepton generations transform under the gauge symmetry as
\begin{equation}
f^{a}_L = \left( \begin{array}{c}
               \nu^{a}_{L}\\ e^{a}_{L}\\ (\nu^{c}_{R})^a\\ (e^{c}_{R})^a
               \end{array}  \right)_{L} \sim (1, 4, 0),
\end{equation}
where a=1,2,3 is a generation index.
In the quark sector, two of the three quark generation transform identically
and one generation transforms in a different
representation of $\mbox{SU}(4)_{L}\otimes \mbox{U}(1)_{N}$.
The quarks have the following representation under the
$\mbox{SU}(3)_C\otimes \mbox{SU}(4)_L\otimes \mbox{U}(1)_N$ gauge group:
\[ Q_{1L} = \left( \begin{array}{c}
                 u_{1}\\ d_{1}\\ u^{'}_1\\ J_{1}
                \end{array}  \right)_{L} \sim (3, 4, 2/3),\]
\[ u_{1R}\sim (3, 1, 2/3), d_{1R}\sim (3,  1, -1/3),
u^{'}_{1R}\sim (3, 1, 2/3), J_{1R}\sim (3, 1, 5/3),\]
\[Q_{iL} = \left( \begin{array}{c}
              d_{i}\\ u_{i}\\ d^{'}_{i}\\J_i\\
                \end{array}  \right)_{L} \sim (3, \bar{4}, -1/3),\]
\begin{equation}
 u_{iR}\sim (3, 1, 2/3), d_{iR}\sim (3, 1, -1/3),
d^{'}_{iR}\sim (3,  1, 2/3), J_{iR}\sim (3, 1, -4/3),
\end{equation}
where $i = 2,3$.
All gauge anomalies cancel in this theory. As discussed in Ref.[3] this type of
construction is only anomaly free when the number of generations is divisible
by 3.
In the fermion representations we have added right-handed neutrinos and
the exotic quarks $u^{'}_1, d^{'}_{2,3}, J_{1,2,3}$.

We now discuss symmetry breaking in this model.
We introduce the Higgs field
\begin{equation}
\chi_1 \sim (1, 4, -1),
\end{equation}
which couples via the Yukawa Lagrangian
\begin{equation}
{\cal L}_{Yuk}^{\chi_1} = \lambda_{1}\bar{Q}_{1L}J_{1R}\chi_1 +
 \lambda_{1ij}\bar{Q}_{iL}J_{jR}\chi_1^{*} + \mbox{H.c.},
\end{equation}
where $i,j\ =\ 2, 3$. If $\chi_1$ gets the VEV:
\begin{equation}
\langle \chi_1 \rangle = \left( \begin{array}{c}
                     0\\ 0\\ 0\\ w_{1}
                     \end{array} \right),
\end{equation}
then the exotic charged 5/3 and -4/3 quarks ($J_{1,2,3}$) gain masses.
In order for $u^{'}_1$ and $d^{'}_{2,3}$ to gain masses, we introduce the
Higgs field
\begin{equation}
\chi_2 \sim (1, 4, 0),
\end{equation}
which couples via the Yukawa Lagrangian
\begin{equation}
{\cal L}_{Yuk}^{\chi_2 } = \lambda_{2}\bar{Q}_{1L}u^{'}_{1R}\chi_2 +
 \lambda_{2ij}\bar{Q}_{iL}d^{'}_{jR}\chi_2^{*} + \mbox{H.c.},
\end{equation}
where $i, j\ =\ 2, 3$ and $\chi_2$ gets the VEV:
\begin{equation}
\langle \chi_2 \rangle = \left( \begin{array}{c}
                    0\\ 0\\ w_{2}\\ 0
                    \end{array} \right).
\end{equation}
With the two Higgs fields $\chi_{1,2}$ the gauge symmetry is broken
to the standard model, as indicated below:
\begin{eqnarray}
\mbox{SU}(3)_{C}\hspace*{-0.2cm}&\otimes
&\hspace*{-0.2cm}\mbox{SU}(4)_{L}\otimes
\mbox{U}(1)_{N}\nonumber\\
                &        &\downarrow \langle \chi_1 \rangle, \langle \chi_2
\rangle            \\
\mbox{SU}(3)_{C}\hspace*{-0.2cm}&\otimes &\hspace*{-0.2cm}\mbox{SU}(2)_{L}
           \otimes
\mbox{U}(1)_{Y}\nonumber
\end{eqnarray}
where $Y$ is the linear combination of $\lambda_8, \lambda_{15}$ and $N$
which annihilates $\langle \chi_1 \rangle$ and $\langle \chi_2 \rangle$
and one can easily check that it is given by:
\begin{equation}
Y = 2N - \frac{1}{\sqrt{3}}\lambda_{8}- \frac{2 \sqrt{6}}{3}\lambda_{15},
\end{equation}
where $\lambda_{8}$ and $\lambda_{15}$ are diagonal SU(4) generators
with $\lambda_{8} = diag(1,1,-2,0)/\sqrt{3}$ and
$\lambda_{15} = diag(1,1,1,-3)/\sqrt{6}$. One can easily check that $Y$ is
numerically identical to the standard model hypercharge.

Electroweak symmetry breaking and the fermion masses are assumed to be due to
the VEVs of the Higgs bosons:
\begin{equation}
\rho \sim (1, 4, 1),\ \eta \sim (1, 4,0),\ S \sim (1, 10, 0)
\end{equation}
These Higgs bosons couple to the fermions through the Yukawa Lagrangian:
\[ {\cal L}_{Yuk}^{\rho} = \lambda_{1a}\bar{Q}_{1L}d_{aR}\rho +
 \lambda_{ia}\bar{Q}_{iL}u_{aR}\rho^{*} + \mbox{H.c.},\]

\[ {\cal L}_{Yuk}^{\eta} = \lambda_{1a}^{'}\bar{Q}_{1L}u_{aR}\eta +
 \lambda_{ia}^{'}\bar{Q}_{iL}d_{aR}\eta^{*} + \mbox{H.c.},\]
\begin{equation}
{\cal L}_{Yuk}^{S} = G_{ab}\bar{f}_{aL}(f_{aR})^{c}S + \mbox{H.c.},
\end{equation}
where $a, b = 1, 2, 3$ and $i, j = 2, 3$. If  $\rho $ gets the VEV:
\begin{equation}
\langle \rho \rangle = \left( \begin{array}{c}
                    0\\ u\\ 0\\0
                    \end{array} \right),
\end{equation}
two up- and one down-quarks gain mass. If $\eta $ gets the VEV:
\begin{equation}
\langle \eta \rangle = \left( \begin{array}{c}
                     v\\ 0\\0 \\0
                    \end{array} \right),
\end{equation}
then the remaining quarks get masses.
If $S$ gets the VEV (note that the 10 representation
of SU(4) can be represented as a $4 \times 4$ symmetric matrix):
\begin{equation}
\langle S \rangle = \left( \begin{array}{cccc}
                    0 &   0                    & 0 & 0\\
                    0 &   0                    & 0 & \frac{v^{'}}{\sqrt{2}}\\
                    0 &   0                    & 0 & 0    \\
                    0 & \frac{v^{'}}{\sqrt{2}} & 0 & 0
                    \end{array} \right),
\end{equation}
then all of the leptons get masses. With the VEVs $\langle \rho \rangle,
\langle \eta \rangle, \langle S\rangle$
the intermediate electroweak gauge symmetry is spontaneously broken as
follows:
\begin{eqnarray}{cc}
&\mbox{SU}(3)_{C}&\hspace*{-0.2cm}\otimes \ \mbox{SU}(2)_{L}\otimes
\mbox{U}(1)_{Y}\nonumber \\
&\downarrow      &\hspace*{-0.8cm}<\rho >, <\eta >, <S>   \\
&\mbox{SU}(3)_{C}&\hspace*{-0.2cm}\otimes \ \mbox{U}(1)_{Q}
\nonumber
\end{eqnarray}
The electric charge operator has been identified as $Q = I_3 + {Y \over 2}$
where $I_3$ is given by the $\mbox{SU}(4)$ generator diag(1,-1,0,0)/2 and
$Y$ is given in eq.(21).

Note that the VEVs of $\rho, \eta$ and $S$ transform
as $Y=1, \ \mbox{SU}(2)_L$ doublets (under the $\mbox{SU}(2)_L \otimes \mbox{U}
(1)_Y$ subgroup of $\mbox{SU}(3)_L \otimes \mbox{U}(1)_N$ which is left
unbroken
by $\langle \chi_{1,2} \rangle$). For this reason and the fact that
$Y$ (eq.(21)) is identical to the standard model hypercharge,
it is clear that the model reduces to the
standard model with the correct low energy phenomenology provided $\langle
\chi_{1,2} \rangle \gg
\langle \rho \rangle, \ \langle \eta \rangle,\ \langle S \rangle$.

Note that in the limit $\langle \chi_2 \rangle \gg \langle \chi_1 \rangle,
\langle \rho \rangle, \langle \eta \rangle, \langle S \rangle$, the model
reduces to the Pissano-Pleitez model (with right-handed singlet neutrinos)
at a energy scale much less than $\langle \chi_1 \rangle$.
On the other hand, if  $\langle \chi_1 \rangle \gg \langle \chi_2 \rangle,
\langle \rho \rangle, \langle \eta \rangle, \langle S \rangle$, the model
reduces to the $\mbox{SU}(3)_C \otimes \mbox{SU}(3)_L \otimes \mbox{U}(1)_N$
model discussed before (in the present paper), which has $\nu_R^a$ in
the $\mbox{SU}(3)_L$ triplets.

Motivated by the Pisano-Pleitez model, we looked for an interesting
extensions which can include right-handed neutrinos.
We found an alternative $\mbox{SU}(3)_C \otimes \mbox{SU}(3)_L \otimes
\mbox{U}(1)_N$
model with right-handed
neutrinos which is actually simpler than the original Pisano-Pleitez model.
We then pointed out the existence of another Pisano-Pleitez
type model with right-handed neutrino, which was had
the gauge symmetry $\mbox{SU}(3)_C \otimes \mbox{SU}(4)_L \otimes
\mbox{U}(1)_N$.\\

\vskip .7cm
\noindent
{\large \bf Acknoledgement}
\vskip .5cm
\noindent
R.F. was supported by the National Science and Engineering Research Council of
Canada, and les Fonds FCAR du Qu\'ebec. T.A.T. would like to thank to
Professors P.C.West and H.Minakata for useful discussions.

\vskip 2cm
\noindent
{\bf References}
\vskip 1cm
\noindent
[1] F. Pisano and V. Pleitez, Phys. Rev. D {\bf 46}, 410 (1992);
 R. Foot, O. F. Hernandez, F. Pisano and V. Pleitez, Phys.Rev. D {\bf 47}, 4158
(1993);
P. H. Frampton, Phys. Rev. Lett. {\bf 69}, 2889 (1992).

\vskip .5cm
\noindent
[2] J. C. Montero, F. Pisano and V. Pleitez, Phys. Rev. D {\bf 47}, 2918
(1993);
V. Pleitez and M. D. Tonasse, Phys. Rev. D {\bf 48}, 2353, 5274 (1993);
T. V. Duong and E. Ma, Phys. Lett. B{\bf 316}, 307 (1993);
F. Pisano and V. Pleitez, IFT-P.026/93 preprint (to be published);
D. Ng, TRIUMF preprint.

\vskip .5cm
\noindent
[3] This gauge group and fermion representations were already discussed
by  F.Pisano and Tran Anh Tuan, [ICTP preprint IC/93/200;
{\it in} Proc. of the XIVV Encontro
National de F\'isica de Part\'icular e Campos, Caxambu, 1993]. However,
they did not consider symmetry breaking, which is done
in the present paper.

\vskip .5cm
\noindent
[4] See for example, P. Langacker and D. London, Phys. Rev. D38, 886 (1988).
\end{document}